\documentclass[%
superscriptaddress,
showpacs,
nofootinbib,
amsmath,amssymb,
aps,
prc,
twocolumn,
floatfix ]%
{revtex4-1}

\usepackage{comment}
\usepackage[dvips]{graphicx}
\usepackage{braket}
\usepackage{cases}
\usepackage{caption}

  \makeatletter
    
    \@addtoreset{equation}{section}
  \makeatother

\begin{document}

\renewcommand{\thefootnote}{\fnsymbol{footnote}}

\title{Low-lying collective excited states in non-integrable pairing models
based on stationary-phase approximation to path integral}%

\author{Fang Ni}
 \affiliation{Faculty of Pure and Applied Sciences,
              University of Tsukuba, Tsukuba 305-8571, Japan}
\author{Nobuo Hinohara}
 \affiliation{Center for Computational Sciences,
              University of Tsukuba, Tsukuba 305-8577, Japan}
  \affiliation{Faculty of Pure and Applied Sciences,
              University of Tsukuba, Tsukuba 305-8571, Japan}
\author{Takashi Nakatsukasa}
 \affiliation{Faculty of Pure and Applied Sciences,
              University of Tsukuba, Tsukuba 305-8571, Japan}
 \affiliation{Center for Computational Sciences,
              University of Tsukuba, Tsukuba 305-8577, Japan}
 \affiliation{iTHES Research Group, RIKEN, Wako 351-0198, Japan}

\begin{abstract}
For a description of large-amplitude collective motion associated with
nuclear pairing, requantization of time-dependent mean-field dynamics
is performed using the stationary-phase approximation (SPA) to the path
integral.
We overcome the difficulty of the SPA, which is known to be applicable to integrable systems only,
by developing a requantization approach combining the SPA with the adiabatic self-consistent collective coordinate method (ASCC+SPA).
We apply the ASCC+SPA to 
multi-level pairing models, which are non-integrable systems,
to study the nuclear pairing dynamics.
The ASCC+SPA gives a reasonable description of low-lying excited $0^+$
states in non-integrable pairing systems.
\end{abstract}

\maketitle

\section{Introduction}
Pairing correlation plays an important role in open-shell nuclei.
Effect of the pairing is prominent in many observables for the ground states,
such as the odd-even mass difference, moment of inertia of rotational bands,
and common quantum number $J^\pi=0^+$ for even-even nuclei \cite{RS80}.
Collective excitations associated with the pairing, such as
pair vibrations and pair rotations,
have been observed in a number of nuclei \cite{BB05}.
Most of these states that are ``excited'' from neighboring even-even systems
are associated with the ground $J^\pi=0^+$ states in even-even nuclei.
In contrast, properties of excited $J^{\pi}=0^+$ states are not clearly
understood yet \cite{HW11, G01}, for which the pairing dynamics plays
an important role in a low-energy region (a few MeV excitation) of nuclei. 
In this paper,
we aim to understand the dynamics associated with pairing correlation in nuclei
from a microscopic view point.

The time-dependent mean-field (TDMF) theory is a standard theory to describe
the dynamics of nuclei from the microscopic degrees of freedom.
Inclusion of the pairing dynamics leads to the
time-dependent Hartree-Fock-Bogoliubov (TDHFB) theory,
which has been utilized for a number of studies of nuclear reaction
and structure \cite{NMMY16}.
The small-amplitude approximation of the TDHFB with modern energy density
functionals, the quasiparticle random-phase approximation (QRPA), 
has successfully reproduced the properties of giant resonances in nuclei.
In contrast, the QRPA description of low-lying quadrupole vibrations
is not as good as that of giant resonances \cite{NMMY16}.
A large-amplitude nature of the quantum shape fluctuation is supposed to
be important for these low-lying collective states.
Five-dimensional collective Hamiltonian (5DCH) approaches
have been developed for studies of low-lying quadrupole states,
in which the collective Hamiltonian is constructed from
microscopic degrees of freedom using the mean-field calculation and
the cranking formula for the inertial masses \cite{Bar11,Del10,Fu13}.
The 5DCH model is able to take into account fluctuations of
the quadrupole shape degrees of freedom which are important in many
nuclear low-energy phenomena, such as
shape coexistence and anharmonic quadrupole vibration.
However, the calculated inertia is often too small to reproduce
experimental data, due to the lack of time-odd components
in the cranking formula \cite{RS80}.
This deficiency can be remedied in the adiabatic self-consistent collective
coordinate (ASCC) method \cite{MNM00},
in which the time-odd effect is properly treated.
In addition, the ASCC method enables us to identify a collective subspace
of interest.
The ASCC was developed from the basic idea of the self-consistent
collective coordinate (SCC) method by Marumori and coworkers \cite{MMSK80}.
It has been applied to the nuclear quadrupole dynamics including
the shape coexistence \cite{KNMM05, HNMM08, HNMM09}.

The TDMF (TDHFB) theory corresponds to an SPA solution in the path integral
formulation \cite{Neg82}.
It lacks a part of quantum fluctuation that is
important in the large-amplitude dynamics.
To introduce the quantum fluctuation based on the TDHFB theory,
the requantization is necessary \cite{Neg82,L80,LNP80,Rei80,KS80,K81}.
A simple and straightforward way of requantization is
the canonical quantization.
This is extensively utilized for collective models in nuclear physics.
For instance,
the canonical quantization of the 5DCH was employed for the study of low-lying
excited states in nuclei \cite{PR09,HSNMM10,NMMY16}.
The similar quantization was also utilized for the pairing collective
Hamiltonian \cite{BBPK70, GPBW85, ZPPRS99, P07}.
In our previous work \cite{NN18},
we studied various requantization methods for the two-level pairing model,
to investigate low-lying excited $0^+$ states.
Since the collectivity is rather low in the pairing motion in nuclei,
the canonical quantization often fails to produce an approximate value 
to the exact solution.
In contrast,
the stationary-phase approximation (SPA) to the path integral \cite{SM88} 
can give quantitative results not only for the excitation energies, 
but also for the wave functions and two-particle-transfer strengths.
The quantized states obtained in the SPA have two advantages;
first, the wave functions are given directly in terms of
the microscopic degrees of freedom, and second,
the restoration of the broken symmetries are automatic.
In the pairing model, the quantized states are eigenstates of the particle-number
operator.
On the other hand, applications of the SPA have been limited to
integrable systems.
This is because we need to find separable periodic trajectories on a classical
torus.
Since the nuclear systems, of course, correspond to non-integrable systems,
a straightforward application of the SPA is not possible.

In this paper,
we propose a new approach of the SPA applicable to the non-integrable systems,
which is based on the extraction of the one-dimensional (1D)
collective coordinate
using the ASCC method.
Since the 1D system is integrable,
the collective subspace can be quantized with the SPA.
The optimal degree of freedom associated with a slow collective motion
is determined self-consistently
inside the TDHFB space, without any assumption. 
Thus, our approach of the ASCC+SPA to the pairing model
basically consists of two steps:
(1) find a decoupled 1D collective coordinate of the pair vibration,
in addition to the pair rotational degrees of freedom.
(2) apply the SPA independently to each collective mode.

The paper is organized as follows. 
Section \ref{sec2} introduces the theoretical framework of the ASCC, SPA,
and their combination, ASCC+SPA. 
In Sec. \ref{sec3}, we provide some details in the application
of the ASCC+SPA to the multi-level pairing model.
We give the numerical results in Sec. \ref{sec4}, including
neutron pair vibrations in Pb isotopes.
The conclusion and future perspectives are given in Sec. \ref{sec5}.


\section{Theoretical framework}
\label{sec2}

\subsection{Adiabatic self-consistent collective coordinate method and
1D collective subspace}
\label{sec:ASCC}
In this section,
we first recapitulate the ASCC method to find a 1D collective coordinate,
following the notation of Ref. \cite{N2012}.

As is seen in Sec.~\ref{sec3}, the TDHFB equations can be
interpreted as the classical Hamilton's equations of motion
with canonical variables $\{\xi^{\alpha},\pi_{\alpha}\}$. 
Each point in the phase space $(\xi^{\alpha},\pi_{\alpha})$
corresponds to a generalized Slater determinant (coherent state).
Assuming slow collective motion,
we expand the Hamiltonian $\mathcal{H}(\xi,\pi)$ with respect to
the momenta $\pi$ up to second order.
The TDHFB Hamiltonian is written as
\begin{align}
 \mathcal{H} &= V(\xi) + \frac{1}{2}B^{\alpha\beta}(\xi)\pi_{\alpha}\pi_{\beta}
\end{align}
with the potential $V(\xi)$ and
the reciprocal mass parameter $B^{\alpha\beta}(\xi)$ defined by
\begin{align}
  V(\xi) &= \mathcal{H}(\xi,\pi=0) , \\
  B^{\alpha\beta}(\xi) &= \left. \frac{\partial^2\mathcal{H}(\xi,\pi)}{\partial\pi_{\alpha}\partial\pi_{\beta}} \right|_{\pi=0}.
\end{align}

For multi-level pairing models in Sec.~\ref{sec3}, 
there is a constant of motion in the TDHFB dynamics,
namely the average particle number $q^n\equiv \langle \hat{N} \rangle/2$.
Since the particle number $\hat{N}$ is time-even Hermitian operator, we treat
this as a coordinate, and its conjugate gauge angle, $p_n$,
as a momentum.
Since $q^n$ is a constant of motion, the Hamiltonian does not
depend on $p_n$.
On the other hand, the gauge angle $p_n$ changes in time,
which corresponds to the pair rotation, a Nambu-Goldstone (NG) mode
associated with the breaking of the gauge (particle-number) symmetry.
We assume the existence of 2D collective subspace $\Sigma_2$
(4D phase space),
described by a set of canonical variable $(q^1,q^2=q^n;p_1,p_2=p_n)$,
which is well decoupled from the rest of degrees of freedom,
$\{q^a,p_a\}$ with $a=3,\cdots$.
The collective Hamiltonian is given by imposing $q^a=p_a=0$,
namely by restricting the space into the collective subspace
\begin{equation}
\mathcal{H}_{\rm coll}(q,p;q^n)
= \bar{V}(q^1,q^n) + \frac{1}{2}\bar{B}^{11}(q^1,q^n)p_1^2 .
  \label{coll}
\end{equation}
Since there exist two conserved quantities, $q^n$ and $\mathcal{H}_{\rm coll}$,
this 2D system is integrable.
We can treat the collective motion of $(q^1,p_1)$ separately from
the pair rotation $(q^n, p_n)$.

In the collective Hamiltonian (\ref{coll}), the variable $q^n$ is
trivially given as the particle number, which is expanded up to
second order in the momenta $\pi$,
\begin{equation}
q^n=\frac{\langle \hat{N} \rangle}{2} = f^n(\xi)
	+\frac{1}{2}f^{(1)n\alpha\beta}\pi_\alpha\pi_\beta .
	\label{q^n}
\end{equation}
To obtain the non-trivial collective variables
$(q^1,p_1)$, we assume the point transformation\footnote{
We may lift the restriction to the point transformation,
as Eq. (\ref{q^n}) \cite{Sat18}.
In this paper, we neglect these higher-order terms, such as
$f^{(1)1\alpha\beta}\pi_\alpha\pi_\beta/2$.
},
\begin{equation}
  q^1 = f^1(\xi) \label{point} ,
\end{equation}
and $\xi^\alpha$ on the subspace $\Sigma_2$ is given as
$\xi^{\alpha} = g^{\alpha}(q^1,q^n,q^a=0)$.
The momenta on $\Sigma_2$ are transformed as
\begin{align}
p_1 &= g_{,1}^{\alpha}\pi_{\alpha} , \quad p_n = g_{,n}^\alpha \pi_\alpha,
	\label{coll_momenta}\\
\pi_{\alpha} &= f^1_{,\alpha}p_1 +f^n_{,\alpha}p_n,
 \label{momenta}
\end{align}
where the comma indicates the partial derivative
$(f^1_{,\alpha}=\partial f^1/\partial \xi^{\alpha})$. 
The Einstein's convention for summation with respect to the repeated
upper and lower indices is assumed hereafter.
The canonical variable condition leads to
\begin{align}
f^i_{,\alpha}g_{,j}^{\alpha} = \delta^i_j,
  \label{canonicity}
\end{align}
where $i,j=1$ and $n$.
The collective potential $\bar{V}(q^1,q^n)$ and
the collective mass parameter
$\bar{B}_{11}(q^1,q^n)=[\bar{B}^{11}(q^1,q^n)]^{-1}$ can be given by
\begin{align}
\bar{V}(q^1,q^n) &= V(\xi=g(q^1,q^n,q^a=0)), \\
\bar{B}^{11}(q^1,q^n) &= f^1_{,\alpha}\tilde{B}^{\alpha\beta}(\xi)f^1_{,\beta} ,
  \label{coll_mass}
\end{align}
where $\tilde{B}^{\alpha\beta}$ are defined as
\begin{align}
 \tilde{B}^{\alpha\beta}(\xi)
&= B^{\alpha\beta}(\xi) - \bar{V}_{,n} f^{(1)n\alpha\beta}(\xi)
\label{tildeB}.
\end{align} 


Decoupling conditions for the collective subspace $\Sigma_2$ lead to
the basic equations of the ASCC method \cite{MNM00,N2012},
which determine tangential vectors,
$f^1_{,\alpha}(\xi)$ and $g_{,1}^{\alpha}(q)$.
\begin{align}
  \delta H_M(\xi,\pi) = 0, \label{mfHFB}
  \end{align}\begin{align}
 \mathcal{M}^\beta_\alpha f^1_{,\beta}  = \omega^2 f^1_{,\alpha},\hspace{2em} 
\mathcal{M}^\beta_\alpha g^{\alpha}_{,1} = \omega^2 g^{\beta}_{,1} .
  \label{mfQRPA}
\end{align}
The first equation (\ref{mfHFB}) is called moving-frame
Hartree-Fock-Bogoliubov (HFB) equation.
The moving-frame Hamiltonian $\mathcal{H}_M$
is
\begin{align}
\mathcal{H}_M(\xi,\pi) &= \mathcal{H}(\xi,\pi)
	-\lambda_{1} q^1(\xi) - \lambda_{n} q^n(\xi,\pi) .
\label{H_M}
\end{align}
The second equation (\ref{mfQRPA}) is called moving-frame QRPA equation.
The matrix $\mathcal{M}^\beta_\alpha$ in the moving-frame QRPA
equation (\ref{mfQRPA}) can be rewritten as 
 \begin{align}
\mathcal{M}^{\beta}_{\alpha} =
	 \tilde{B}^{\beta\gamma}
	 \left(V_{,\gamma\alpha}-\lambda_{n}f^n_{,\gamma\alpha}\right)
	+ \frac{1}{2}\tilde{B}^{\beta\gamma}_{,\alpha}V_{,\gamma} 
\label{M}.
\end{align}
The NG mode, $f^n_{,\alpha}$ and $g^\alpha_{,n}$, corresponds to
the zero mode with $\omega^2=0$.
Therefore, the collective mode of our interest corresponds to
the mode with the lowest frequency squared except for the zero mode.

In practice, we obtain the collective path according to the following procedure:
\begin{enumerate}
\item Find the HFB minimum point $\xi^\alpha_i$ ($i=0$)
	by solving Eq. (\ref{mfHFB}) with $\lambda_1=0$.
	Let us assume that this corresponds to $q^1_i=0$.
\item \label{step-2}
	Diagonalize the matrix $\mathcal{M}^{\beta}_{\alpha}$
to solve Eq.~(\ref{mfQRPA}) using Eq.~(\ref{M}).
\item \label{step-3}
Move to the next neighboring point
$\xi^\alpha_{i+1}=\xi^\alpha_{i}+d\xi^\alpha$
with $d\xi^{\alpha}=g^{\alpha}_{,1}dq^1$.
This corresponds to the collective coordinate, $q^1_{i+1}=q^1_i+dq^1$.
\item \label{step-4}
At $\xi_{i+1}^\alpha$ $(q^1_{i+1})$,
obtain a self-consistent solution of
Eqs. (\ref{mfHFB}) and (\ref{mfQRPA})
to determine
$\xi^\alpha_{i+1}$, $f^1_{,\alpha}$, and $g^\alpha_{,1}$.
\item Go back to Step \ref{step-3} to determine the next point
on the collective path.
\end{enumerate}
We repeat this procedure with $dq^1 > 0$ and $dq^1<0$, and
construct the collective path.
In Steps~\ref{step-2} and \ref{step-4}, we choose a mode
with the lowest frequency squared $\omega^2$.
Note that $\omega^2$ can be negative.
In Step \ref{step-4}, when we solve Eq.~(\ref{mfHFB}), we use
a constraint on the magnitude of $dq^1=f^1_{,\alpha}d\xi^{\alpha}$.
Since the normalization of $f^1_{,\alpha}$ and $g^\alpha_{,1}$ is
arbitrary as far as they satisfy Eq. (\ref{canonicity}),
we fix this scale by an additional condition of $\bar{B}^{11}(q^1)=1$.


\subsection{Stationary-phase approximation to path integral}
\label{sec:SPA}

For quantization of integrable systems,
we can apply the stationary-phase approximation (SPA)
to the path integral.
In our former study \cite{NN18},
we have proposed and tested the SPA for an integrable pairing model.
Since the collective Hamiltonian (\ref{coll}) that is
extracted from the TDHFB degrees of freedom is integrable,
the SPA is applicable to it.
In this manner, we may apply the ASCC+SPA to
non-integrable systems in general.

\subsubsection{Basic idea of ASCC+SPA}

Because the Hamiltonian $\mathcal{H}_{\rm coll}$ of Eq.~(\ref{coll})
is separable,
it is easy to find periodic trajectories on invariant tori.
Since the pair rotation corresponds to the motion of $p_n$
with a constant $q^n$,
all we need to do is to find classical periodic trajectories $C_k$ in
the $(q^1,p_1)$ space (with a fixed $q^n$) which satisfy
the Einstein-Brillouin-Keller (EBK) quantization rule
with a unit of $\hbar=1$,
\begin{equation}
	\oint_{C_k} p_1dq^1 = 2\pi k ,
	\label{EBK}
\end{equation}
where $k$ is an integer number.

At each point in the space $(q^1,q^n; p_1,p_n)$ corresponds to
a generalized Slater determinant
$\ket{q^1,q^n;p_1,p_n}=\ket{\xi,\pi}$
where $(\xi,\pi)$ are given as
$\xi^\alpha=g^\alpha(q^1,q^n,q^a=0)$ and
$\pi_\alpha=f^1_{,\alpha}p_1+f^n_{,\alpha}p_n$.
According to the SPA, the $k$-th excited state $\ket{\psi_k}$
is constructed from the $k$-th periodic trajectory $C_k$,
given by $(q^1(t),p_1(t))$,
of the Hamiltonian $\mathcal{H}_{\rm coll}$.
\begin{equation}
\ket{\psi_k} \propto \oint dp_n \oint_{C_k} \rho(q,p) dt \ket{q,p}
	e^{i\mathcal{T}[q,p]} ,
	\label{SPA}
\end{equation}
where $(q,p)$ means $(q^1,q^n;p_1, p_n)$, and the weight function
$\rho(q,p)$ is given through an invariant measure $d\mu(q,p)$ as
\begin{equation}
d\mu(q,p)=\rho(q,p) dE dt dq^n dp_n .
\end{equation}
The invariant measure $d\mu(q,p)$ is defined by the unity condition
$\int d\mu(q,p) \ket{q,p}\bra{q,p} = 1$.
An explicit form of $d\mu(q,p)$ for the present pairing model is shown
in Eq. (\ref{dmu}).
The action integral $\mathcal{T}$ is defined by
\begin{align}
\mathcal{T}[q,p] &\equiv
\int_{0}^{t} \braket{q(t'),p(t')| i\frac{\partial}{\partial t'}
	|q(t'),p(t')} dt' .
\label{tau}
\end{align}

The SPA quantization is able to provide a wave function $\ket{\psi_k}$
in microscopic degrees of freedom, which is given as a superposition
of generalized Slater determinants $\ket{q,p}$.
In addition, the integration with respect to $p_n$ over a circuit on a torus
automatically recovers the broken symmetry, namely the good particle number.
However, it relies on the existence of invariant tori.
In the present approach of the ASCC+SPA,
we first derive a decoupled collective subspace $\Sigma_2$ and identify
canonical variables $(q,p)$.
Because of the cyclic nature of $(q^n,p_n)$, it is basically a 1D system
and becomes integrable.
In other words, we perform the torus quantization on
approximate tori in the TDHFB phase space $(\xi,\pi)$,
which is mapped from tori in the 2D collective subspace $(q,p)$.


\subsubsection{Notation and practical procedure for quantization}
\label{sec:notation}

For the application of the ASCC+SPA method to the pairing model in Sec.~\ref{sec3},
we summarize some notations and procedures to obtain quantized states.

In Sec.~\ref{sec3}, the time-dependent generalized Slater determinants
(coherent states) are written as $\ket{Z}$ with complex variables
$Z_\alpha(t)$.
The variables $Z_\alpha$ are transformed into real variables
$(j^\alpha, -\chi_\alpha)$ that correspond to $(\xi^\alpha,\pi_\alpha)$
in Sec.~\ref{sec2}.
$\chi_\alpha$ and $j^\alpha$ correspond to
the ``angle'' and the ``number'' variables, respectively.
Although it is customary to take the time-odd angle $\chi_\alpha$ as a coordinate,
we take the number $j^\alpha$ as a coordinate and the angle $-\chi_\alpha$ as a momentum with an additional minus sign.
Similarly, the gauge angle $\Phi$ and the total particle number $J$
correspond to variables of the pair rotation, $-p_n$ and $q^n$, respectively.


According to the EBK quantization rule (\ref{EBK}),
the ground state with $k=0$ corresponds to nothing but the HFB state
with a fixed particle number $J(=q^n)$.
For the $k$-th excited states, we perform the following calculations:
\begin{enumerate}
\item Obtain the 1D collective subspace with canonical variables $(q^1,p_1)$
	according to the ASCC in Sec.~\ref{sec:ASCC}.
\item Find a trajectory $(q^1(t),p_1(t))$ which satisfies
the $k$-th EBK quantization condition (\ref{EBK}).
\item Calculate the action integral (\ref{tau}) for the $k$-th trajectory.
\item Construct the $k$-th excited state using Eq.~(\ref{SPA}).
\end{enumerate}
The ASCC provides the 2D collective subspace $(q^1,J)$
and the generalized coherent states $\ket{\Phi=0,J;q,p=0}$.
For finite values of momenta, we use Eq.~(\ref{momenta})
to obtain the state $\ket{\Phi,J;q,p}$.


\section{Pairing model}
\label{sec3}

We study the low-lying excited $0^+$ states
in the multi-level pairing model by applying the ASCC+SPA.
The Hamiltonian of the pairing model is given in terms of
the single-particle energies $\epsilon_l$ and the pairing strength $g$ as
\begin{align}
	\hat{H} &= \sum_\alpha \epsilon_\alpha \hat{n}_\alpha - g \sum_{\alpha,\beta} \hat{S}_\alpha^+ \hat{S}_{\beta}^- \nonumber \\
    &= \sum_\alpha\epsilon_\alpha(2\hat{S}_\alpha^0+\Omega_\alpha) - g \hat{S}^+ \hat{S}^{-} ,
\end{align}
where we use the SU(2) quasi-spin operators,
$\boldsymbol{\hat{S}}=\sum_\alpha \boldsymbol{\hat{S}}_\alpha$, with
\begin{eqnarray}
        \hat{S}_\alpha^0 &=& \frac{1}{2}\left(\sum_m\hat{a}_{j_\alpha m}^{\dag}\hat{a}_{j_\alpha m}-\Omega_\alpha\right) ,\\
        \hat{S}_\alpha^{+} &=& \sum_{m>0}\hat{a}_{j_\alpha m}^{\dag}\hat{a}_{j_\alpha\overline{m}}^{\dag} ,
\quad   \hat{S}_\alpha^{-} = \hat{S}_\alpha^{+\dag} .
\end{eqnarray}
Each single-particle energy $\epsilon_\alpha$ possesses $2\Omega_\alpha$-fold
degeneracy ($\Omega_\alpha=j_\alpha+1/2$)
and $\sum_{m>0}$ indicates the summation over $m=1/2,3/2,\cdots,$
and $\Omega_\alpha-1/2$.
The occupation number of each level $\alpha$ is given by
$\hat{n}_\alpha = \sum_m \hat{a}^{\dag}_{j_\alpha m}\hat{a}_{j_\alpha m}
=2\hat{S}_\alpha^0+\Omega_\alpha
$.
The quasi-spin operators satisfy the commutation relations
\begin{equation}
  [\hat{S}_\alpha^0,\hat{S}_\beta^{\pm}] = \pm\delta_{\alpha\beta}\hat{S}_{\alpha}^{\pm},
\quad [\hat{S}_{\alpha}^{+},\hat{S}_{\beta}^{-}] = 2\delta_{\alpha\beta}\hat{S}_{\alpha}^{0} .
\end{equation}
The magnitude of quasi-spin for each level is given by
$S_\alpha=\frac{1}{2}(\Omega_\alpha-\nu_\alpha)$, where $\nu_\alpha$
is the seniority
quantum number, namely the number of unpaired particles at the level $\alpha$.
In the present study, we consider only seniority-zero states with
$\nu=\sum_\alpha \nu_\alpha=0$.
The residual two-body interaction consists of the monopole pairing
interaction which couples two particles to zero angular momentum.
We obtain the exact solutions either by solving the Richardson equation
\cite{Richardson,Richardson2,Richardson3} or
by diagonalizing the Hamiltonian using the quasi-spin symmetry.

\subsection{Classical form of TDHFB Hamiltonian}

The time-dependent coherent state for the seniority $\nu=0$ states
($S_\alpha=\Omega_\alpha/2$) is constructed with
time-dependent complex variables $Z_\alpha(t)$ as
\begin{equation}
	\ket{Z(t)} = \prod_{\alpha} \left(1+|Z_\alpha(t)|^2\right)^{-\Omega_\alpha/2}
	\exp [Z_\alpha(t) \hat{S}_\alpha^{+}] \ket{0},
 \label{coherent}
\end{equation}
where $\ket{0}$ is the vacuum (zero-particle) state.
The TDHFB motion is given by the time development of $Z_\alpha(t)$.
In the SU(2) quasi-spin representation,
$\ket{0}=\prod_\alpha \ket{S_\alpha,-S_\alpha}$.
The coherent state $\ket{Z(t)}$ is a superposition of
the states with different particle numbers
without unpaired particles.
In the present pairing model,
the coherent state is the same as the time-dependent BCS wave function
with $Z_\alpha(t)=v_\alpha(t)/u_\alpha(t)$,
where $(u_\alpha(t),v_\alpha(t))$ are the time-dependent BCS $u,v$ factors.

The TDHFB equation can be derived from the time-dependent variational
principle, $\delta \mathcal{S} = 0$, where
\begin{equation}
	\mathcal{S}\equiv \int \mathcal{L}(t) dt =
	\int \braket{Z(t)|i\frac{\partial}{\partial t}-\hat{H}|Z(t)}dt
  \label{TDHFB}
\end{equation}
($\hbar=1$ applies hereafter).
After the transformation of the complex variables into the real ones with
$Z_\alpha = \tan{\frac{\theta_\alpha}{2}}e^{-i\chi_\alpha}$
($0\leq\theta_\alpha\leq\pi$),
the Lagrangian $\mathcal{L}$ and the expectation value of the Hamiltonian
are written as
\begin{equation}
\mathcal{L}(t) = \sum_\alpha \frac{\Omega_\alpha}{2}
	(1-\cos{\theta}_\alpha)\dot{\chi}_\alpha - \mathcal{H}(Z,Z^*) ,
\end{equation}
with
\begin{align}
	\mathcal{H}( &Z,Z^*) \equiv \braket{Z|\hat{H}|Z}& \nonumber \\
	=& \sum_\alpha \epsilon_\alpha\Omega_\alpha(1- \cos{\theta}_\alpha)
	\nonumber \\
	&- \frac{g}{4}\sum_\alpha \Omega_\alpha [\Omega_\alpha(1-\cos^2{\theta}_\alpha)+(1-\cos{\theta}_\alpha)^2] \nonumber \\
& - \frac{g}{4}\sum_{\alpha\neq \beta} \Omega_{\alpha}\Omega_{\beta} \sqrt{(1-\cos^2{\theta}_{\alpha})(1-\cos^2{\theta}_{\beta})}e^{-i(\chi_{\alpha}-\chi_{\beta})}   .
\label{TDHFB_Hamiltonian_2}
\end{align}
We choose $\chi_\alpha$ as canonical coordinates, and their conjugate momenta
are given by
\begin{align}
  j^\alpha\equiv
	\frac{\partial\mathcal{L}}{\partial\dot{\chi}_\alpha}=\frac{\Omega_\alpha}{2}
	(1-\cos{\theta}_\alpha) .
\end{align}
$\chi_\alpha$ represents a kind of gauge angle of each level,
and $j^\alpha$ corresponds to the occupation number of each level,
$2j^\alpha=\bra{Z} \hat{n}_\alpha\ket{Z}$. 
As we mention in Sec.~\ref{sec:notation},
we switch the coordinates and momenta,
$(\chi_\alpha,j^\alpha) \rightarrow (j^\alpha,-\chi_\alpha)$,
to make the coordinates time even.
The TDHFB equation is equivalent to classical Hamilton's equations
\begin{equation}
	-\dot{\chi}_\alpha = -\frac{\partial\mathcal{H}}{\partial j^\alpha}, \quad
	\dot{j}^\alpha = \frac{\partial\mathcal{H}}{\partial (-\chi_\alpha)} .
\end{equation}

\subsection{Application of ASCC}

We construct a 2D collective subspace $\Sigma_2$ from the ASCC method.
We expand the classical Hamiltonian up to second order with respect to
the momenta, $-\chi_{\alpha}$
\begin{align}
  \mathcal{H}(j,\chi) \approx& V(j) + \frac{1}{2}B^{\alpha\beta}(j)\chi_{\alpha}\chi_{\beta},
\end{align}
where the potential $V(j)$ and the reciprocal mass parameter
$B^{\alpha\beta}(j)$ are given as
\begin{align}
  V(j) =& \mathcal{H}(j,\chi=0) \nonumber \\
	=& \sum_{\alpha} 2\epsilon_{\alpha}j^{\alpha} - g\sum_{\alpha} \left[ \Omega_{\alpha}j^{\alpha} - (j^{\alpha})^2 +\frac{(j^{\alpha})^2}{\Omega_{\alpha}} \right] \nonumber \\
  &- g\sum_{\alpha\ne \beta} \sqrt{j^{\alpha}j^{\beta}(\Omega_{\alpha}-j^{\alpha})(\Omega_{\beta}-j^{\beta})}, 	
\end{align}
\begin{eqnarray}
&&B^{\alpha\beta}(j) = \left. \frac{\partial^2\mathcal{H}}{\partial\chi_{\alpha}\partial\chi_{\beta}} \right|_{\chi=0} \\
\label{mass}
&&=
	\begin{cases}
2g\sum_{\gamma\ne \alpha} \sqrt{j^{\gamma}j^{\alpha}(\Omega_{\gamma}-j^{\gamma})(\Omega_{\alpha}-j^{\alpha})}
		& \text{for $\alpha=\beta$} \\
-2g\sqrt{j^{\alpha}j^{\beta}(\Omega_{\alpha}-j^{\alpha})(\Omega_{\beta}-j^{\beta})}
		& \text{for $\alpha\ne\beta$}
	\end{cases}. \nonumber
\end{eqnarray}
We may apply the ASCC method in Sec. \ref{sec:ASCC}
by regarding $\xi\to j$ and $\pi\to-\chi$.

The TDHFB conserves the average total particle number $N$.
We adopt 
\begin{equation}
	J\equiv N/2=\sum_\alpha j^\alpha, 
  \label{J}
\end{equation}
as a coordinate $q^n$.
Since this is explicitly given as the expectation value of the particle-number
operator, curvature quantities,
such as $f^n_{,\alpha\beta}$ and $f^{(1)n\alpha\beta}$,
are explicitly calculable.
On the other hand, the gauge angle $\Phi=-p_n$ is not given a priori.
Since the ASCC solution provides $g^\alpha_{,n}$ as an eigenvector of
Eq. (\ref{mfQRPA}),
we may construct it as Eq. (\ref{coll_momenta}) in the first order
in $\pi=-\chi$.
We confirm that the pair rotation corresponds to an eigenvector
of Eq. (\ref{mfQRPA}) with the zero frequency $\omega^2=0$.


In the present pairing model, 
from Eq. (\ref{J}), we find $J$ does not depend on $\chi$.
This means $f^{(1)n\alpha\beta}=0$ in Eq. (\ref{q^n}),
thus, $\tilde{B}^{\alpha\beta}=B^{\alpha\beta}$.
The second derivative of $J$ with respect to $j$ also vanishes,
which indicates $f^n_{,\gamma\alpha}$ in Eq.~(\ref{M}) is zero. 
The gauge angle $\Phi$ is locally determined
by the solution of Eq. (\ref{mfQRPA}).
\begin{align}
\Phi = g^\alpha_{,n} \chi_{\alpha},
  \label{total_gauge}
\end{align}


It should be noted \cite{N2012} that the definition of the collective
variables $(q^1,p_1)$ is not unique, because it can be arbitrarily mixed
with the pair rotation $(q^n,p_n)$ as
\begin{equation}
	q^1 \rightarrow q^1 + c q^n,  \quad
	p_n \rightarrow p_n -c p_1 \label{eq:pairrotmix}
\end{equation}
with an arbitrary constant $c$.
Numerically, this arbitrariness sometimes leads to a problematic behavior 
in iterative procedure of the ASCC.
In order to fix the parameter $c$, we adopt a condition
called ``ETOP'' in Ref.~\cite{HNMM07}.
We require the following condition to determine $c$:
\begin{align}
\sum_\alpha 
	f^1_{,\alpha} = 0 ,
\end{align}
where $f^1_{,\alpha}$ is replaced as
\begin{align}
{f}^1_{,\alpha} \to f^1_{,\alpha} + c f^n_{,\alpha}
  \label{f}
\end{align}
with Eq. (\ref{eq:pairrotmix}).

\subsection{Application of SPA}

After deriving the collective subspace $\Sigma_2$, 
we perform the quantization according to the SPA in Sec.~\ref{sec:SPA}.
Calculating a trajectory in the $(q^1,p_1)$ space,
we can identify a series of states $\{\ket{\Phi,J;q^1(t),p_1(t)}\}$
on the trajectory,
in the form of Eq. (\ref{coherent})
with parameters $Z_\alpha$ given at $(\Phi,J,q^1,p_1)$ and $q^a=p_a=0$ for
$a\geq 3$.
Since the variables $(\Phi,J)$ and $(q^1,p_1)$ are separable,
we may take closed trajectories independently in $(\Phi,J)$ and $(q^1,p_1)$
sectors, which we denote here as $C_\Phi$ and $C_1$, respectively.
The action integral is given by
\begin{align}
\mathcal{T}(\Phi,J;q^1,p_1)
&= \int_{C_\Phi} \braket{\Phi(t),J;q^1,p_1|i\frac{\partial}{\partial t}|\Phi(t),J;q^1,p_1} dt \nonumber \\
+ \int_{C_1} &\braket{\Phi,J;q^1(t),p_1(t)|i\frac{\partial}{\partial t}|\Phi,J;q^1(t),p_1(t)} dt
 \nonumber \\
	&=J\Phi + \int_{C_1} \sum_\alpha j^\alpha d\chi_\alpha
	\nonumber \\
	&\equiv \mathcal{T}_\Phi(J,\Phi) +\mathcal{T}_1(q^1,p_1;J) .
\end{align}
In fact, the gauge-angle dependence is formally given as
\begin{equation}
  \ket{\Phi,J;q^1,p_1} = e^{-i\Phi \hat{N}/2} \ket{J;q^1,p_1},
\end{equation}
where $\hat{N}=\sum_\alpha \hat{n}_\alpha$.
Then, the action for the trajectory $C_1$ can be also expressed as
$\mathcal{T}(q^1,p_1;J)=
 \int_{C_1} \braket{J;q^1,p_1|i\frac{\partial}{\partial t}|J;q^1,p_1} dt$.

In the SU(2) representation, the invariant measure is
\begin{align}
  d\mu(Z) &= \prod_\alpha \frac{\Omega_\alpha+1}{\pi} (1+|Z_\alpha|^2)^{-2} d({\rm Re}\,Z) d({\rm Im}\,Z) \nonumber \\
=& \prod_\alpha \frac{-(\Omega_\alpha+1)}{4\pi} d(\cos{\theta_\alpha}) d\chi_\alpha \nonumber \\
 =& \prod_\alpha \frac{1+\Omega_\alpha^{-1}}{2\pi} d\chi_\alpha dj^\alpha \nonumber \\
 =& \left[ \prod_\alpha \frac{1+\Omega_\alpha^{-1}}{2\pi} \right] d\Phi dJ dq^1 dp_1 
  \prod_a dq^a dp_a,
	\label{dmu}
\end{align}
where $(q^a,p_a)$ are the intrinsic canonical variables
decoupled from the collective subspace $\Sigma_2$.
In the last line in Eq. (\ref{dmu}),
we used the invariance of the phase-space volume element
in the canonical transformation.
According to Eq. (\ref{dmu}),
the weight function $\rho(q,p)$ in Eq. (\ref{SPA}) is just a constant,
thus, treated as the normalization of the wave function.

The coherent state $\ket{\Phi,J;q^1,p_1}=\ket{Z}$ is
expanded in the SU(2) quasispin basis as
\begin{align}
\ket{Z} &= 
\sum_{\{m_\alpha\}} A_m(Z) \ket{\cdots;S_\alpha,-S_\alpha+m_\alpha,\cdots}, \nonumber\\
\end{align}
where the summation is taken over all possible combinations
of integer values of $\{ m_\alpha\}$ with
\begin{align}
A_m(Z) =& \prod_\alpha 
\frac{Z_\alpha^{m_\alpha}}
	{\left(1+|Z_\alpha|^2\right)^{\Omega_\alpha/2}m_\alpha!}
\sqrt{\frac{\Omega_\alpha! m_\alpha!}{(\Omega_\alpha-m_\alpha)!}} \nonumber\\
 =& \prod_\alpha \left(\frac{1-\cos{\theta}_\alpha}{2}\right)^{m_\alpha/2}\left(\frac{1+\cos{\theta}_\alpha}{2}\right)^{(\Omega_\alpha-m_\alpha)/2}
  \nonumber \\
  &\times\sqrt{\frac{\Omega_\alpha!}{m_\alpha!(\Omega_\alpha-m_\alpha)!}} e^{-i m_\alpha\chi_\alpha}  ,
	\label{A_m}
\end{align}
where the lower index $m$ indicates a combination of $\{ m_\alpha \}$.
The integer number $m_\alpha$ corresponds to the number of pairs
in the level $\alpha$.
%

Using Eq.~(\ref{A_m}),
the $k$-th excited state is calculated as
\begin{align}
\ket{\psi_k} \propto& \oint_{C_\Phi} d\Phi \oint_{C_1} dt
\ket{\Phi,J;q^1,p_1} e^{i\mathcal{T}(\Phi,J;q^1,p_1)}
 \nonumber \\
	=& \sum_{\{m_\alpha\}}
	\int_0^{2\pi} d\Phi e^{i(J-\sum_\alpha m_\alpha)\Phi} \nonumber \\
&\times \oint dt e^{i\mathcal{T}_1}(t) B_m (Z) \ket{\cdots;S_\alpha,-S_\alpha+m_\alpha,\cdots} \nonumber \\
 \equiv& \sum_{\{m_\alpha\}_J} C_m \ket{\cdots;S_\alpha,-S_\alpha+m_\alpha,\cdots},
 \label{SPA2}
\end{align}
where $B_m(Z)$ are identical to $A_m$ in Eq.~(\ref{A_m}) except for
replacing $\chi_\alpha$ with the relative angles
$\phi_\alpha\equiv\chi_\alpha-\Phi$.
The coefficients $C_m$ are given by
\begin{align}
C_m = \oint_{C_1} dt e^{i\mathcal{T}_1(t)} B_m(Z(t)).
  \label{coef}
\end{align}
In the last line of Eq. (\ref{SPA2}), the summation is restricted to
$\{m_\alpha\}$ that satisfy $\sum_\alpha m_\alpha=J$.
It is easy to find that $J$ must be integer,
according to the quantization rule (\ref{EBK}) for the $(J,\Phi)$ sector.

The SPA for the ground state ($k=0$) is given by the stationary point
in the $(q^1,p_1)$ sector,
namely, the HFB state $\ket{\Phi,J;q,p}=e^{-i\Phi\hat{N}/2}\ket{\rm HFB}$.
Nevertheless, the rotational motion in $\Phi(t)$ is present,
which leads to the number quantization (projection).
Therefore, Eq.~(\ref{SPA2}) becomes
\begin{align}
	\ket{\psi_{\rm g.s.}} \propto& \sum_{\{m_\alpha\}}
	\int_0^{2\pi} d\Phi e^{i(J-\sum_\alpha m_\alpha)\Phi}
	\ket{\rm HFB},  
	\label{SPA3}
\end{align}
which is identical to the wave function of the particle-number projected HFB state.


\section{Results}
\label{sec4}

In the pairing model in Sec.~\ref{sec3},
the number of TDHFB degrees of freedom equals that of single-particle levels.
As there are two constants of motion, that is, the particle number and 
the energy, the system is integrable for one- and two-level models.
We first apply the ASCC+SPA method to an integrable two-level model,
then, to non-integrable multi-level models.

\subsection{Integrable case: Two-level pairing model}

\begin{figure}[thb]
 \begin{center}
   \includegraphics[height=0.45\textwidth,angle=-90]{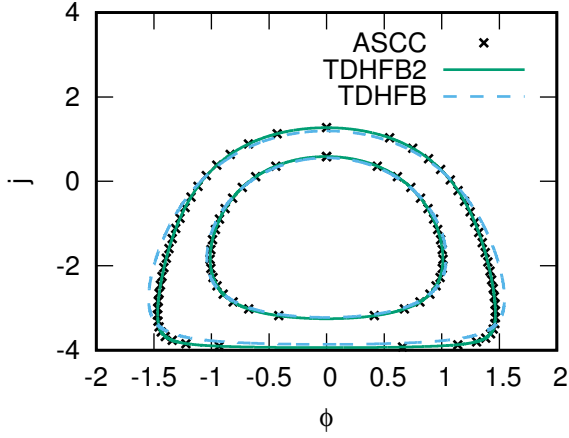}
 \end{center}
\caption{Classical trajectories satisfying the EBK quantization condition
(\ref{EBK}) with $k=1$ and 2 in the $(\phi,j)$ phase space.
The crosses, solid and dashed lines correspond to the results
of the ASCC+SPA, TDHFB2+SPA, and TDHFB+SPA, respectively.
The crosses for the ASCC+SPA trajectories are plotted every ten calculations
($\delta q = 10dq =0.1/\sqrt{\epsilon_0}$). 
}
\label{fig:N16_traj}
\end{figure}
\begin{figure}[thb]
 \begin{center}
   \includegraphics[height=0.5\textwidth,angle=-90]{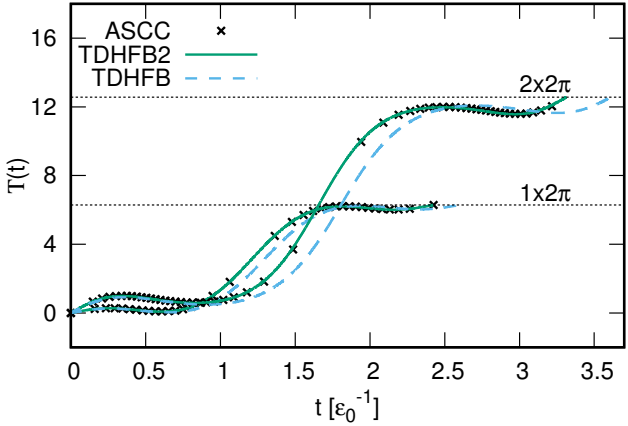}
 \end{center}
\caption{Calculated action integrals for the $\ket{0_2^+}$ and $\ket{0_3^+}$ states as functions of time $t$.
The crosses, solid and dashed lines correspond to the ASCC+SPA,
the TDHFB2+SPA, and the TDHFB+SPA, respectively. 
The action integrals are calculated on each trajectory in Fig. \ref{fig:N16_traj}
from $(\phi,j)=(0,j_{\rm max})$ in the clockwise direction.
The crosses for the ASCC+SPA trajectories are plotted every ten calculations
($\delta q = 10dq =0.1/\sqrt{\epsilon_0}$). 
}
\label{fig:N16_tau}
\end{figure}
\begin{figure}[thb]
 \begin{center}
   \includegraphics[height=0.5\textwidth,angle=-90]{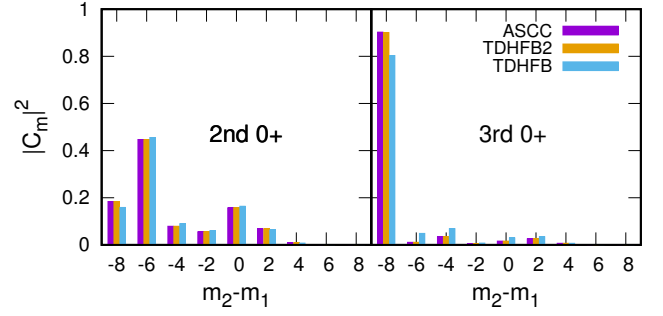}
 \end{center}
\caption{
Occupation probabilities for the $0_2^+$ and $0_3^+$ states.
The horizontal line indicates the $j=(m_2-m_1)/2$ of the quasi-spin basis
in Eq. (\ref{SPA2}).
The vertical bars at each $m_2-m_1$ from left to right
represent $|C_m|^2$ of Eq. (\ref{coef}) in
the ASCC+SPA, TDHFB2+SPA and TDHFB+SPA calculations,
respectively.
}
 \label{fig:N16_occ}
\end{figure}

The two-level pairing model corresponds to the 2D TDHFB system.
Explicitly separating the gauge angle $\Phi$ and
fixing the particle number $J$,
the 2D TDHFB is reduced to the 1D system
with the relative angle $\phi\equiv \chi_2-\chi_1$ and the relative
occupation $j\equiv (j_2-j_1)/2$ as canonical variables.
In Ref.~\cite{NN18}, using the explicit transformation to these
separable variables,
we examined the performance of the SPA requantization
for the two-level model.
In this section, we discuss the same model as in Ref.~\cite{NN18},
but we determine the transformation using the ASCC method and then apply the
SPA (ASCC+SPA).

Here, we study the system with the equal degeneracy,
$\Omega_1=\Omega_2=8$, 
the pairing strength $g/\epsilon_0=0.2$,
and the particle number $N=16$. 
In this two-level case, we use the level spacing,
$\epsilon_0\equiv \epsilon_2-\epsilon_1$,
as the unit of energy.
The moving-frame QRPA produces the zero mode and another eigenvector
with a finite frequency squared $\omega^2\neq 0$.
We follow the latter mode to construct the collective path.
In the ASCC calculation, we set the increment of the
collective coordinate, $dq=0.01$,
in units of $1/\sqrt{\epsilon_0}$.
We confirm that the pair rotation always has a zero frequency
on the collective path.
On the obtained collective path, we calculate a classical trajectory
for the Hamiltonian
\begin{equation}
\mathcal{H}_{\rm coll}(q^1,p_1;J)=\frac{1}{2} p_1^2 + V(q^1,J)
\end{equation}
with $J=N/2=8$.
Calculated trajectories that satisfy the EBK quantization condition
(\ref{EBK}) for the first and second excited states ($0_2^+$ and $0_3^+$)
are mapped onto the $(\phi,j)$ plane and
shown in Fig. \ref{fig:N16_traj}.
We also calculate the trajectories using the explicit transformation
of the variables to $(\Phi,J;\phi,j)$, which
are shown by dashed lines in Fig. \ref{fig:N16_traj}.
We call this ``TDHFB trajectories''.
Small deviation in large $\phi$ is due to the absence of the higher-order
terms in $\chi_\alpha$ in the ASCC.
In fact, if we calculate the trajectories in the variables
$(\Phi,J;\phi,j)$ using the Hamiltonian truncated up to the
second order in $\chi_\alpha$ (``TDHFB2 trajectories''),
we obtain the solid lines in Fig. \ref{fig:N16_traj},
which perfectly agree with the ASCC trajectories.

The action integrals $\mathcal{T}(t)$ corresponding to
these closed trajectories are shown in Fig. \ref{fig:N16_tau}.
For the $0_2^+$ state,
all three calculations agree well
with each other, while we see small deviation between the full TDHFB
and the ASCC/TDHFB2 calculations for the $0_3^+$ state.

The calculated wave functions for the excited $0^+$ states 
are shown in Fig. \ref{fig:N16_occ}. 
We show the occupation probabilities which are decomposed 
into the $2n$-particle-$2n$-hole components. 
The left end of the horizontal axis at $m_2-m_1=-8$ $(j=-4)$
corresponds to a state with $(m_1,m_2)=(N/2,0)$ where all the particles
are in the lower level $\alpha=1$.
The next one at $m_2-m_1=-6$ $(j=-3)$ corresponds to the one with
$(m_1,m_2)=((N-2)/2,1)$, and so on.
The results from the TDHFB2+SPA and the ASCC+SPA are identical to each other
within numerical error, and they reproduce
the TDHFB+SPA calculation well.

%
%

By comparing with the full TDHFB calculation with the ASCC+SPA approach
in the two-level pairing model,
we conclude that the ASCC is reliable for description of low-lying
collective states, for which the adiabatic approximation is justifiable.
In addition to that, we should note that the pair rotation is properly separated.

\begin{figure*}[tb]
 \begin{center}
  \includegraphics[width=40mm,angle=-90]{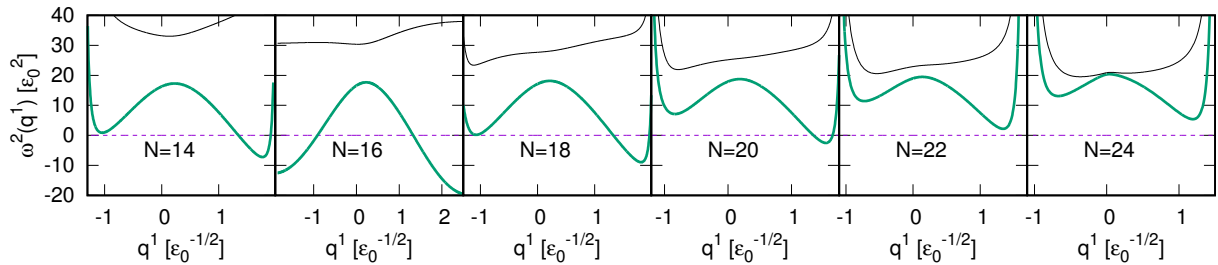}
 \end{center}
 \caption{Eigenvalues of moving-frame QRPA equation as a function of
the collective coordinate $q^1$, from $N=14$ to $N=24$.
The thick (green) lines are the modes we choose as
the collective coordinate $q^1$,
while the dashed lines correspond to the zero modes ($q^n$).
In each panel, both ends of the horizontal axis corresponds
to the ending points of the collective path $q^1$.
}
 \label{omega_sq}
\end{figure*}
\begin{figure*}[tb]
 \begin{center}
  \includegraphics[width=40mm,angle=-90]{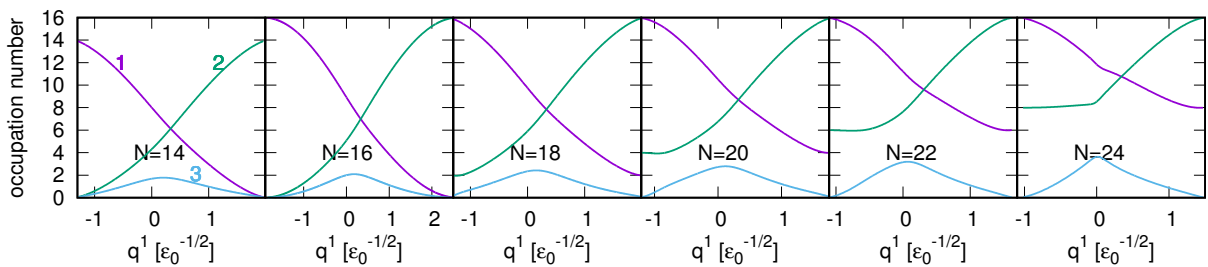}
 \end{center}
\caption{The occupation numbers $2j^{\alpha}$ in each single-particle level $\alpha$
as a function of the collective coordinate $q^1$,
from $N=14$ to $N=24$.
The purple, green, and blue lines correspond to
$\alpha=1$, 2, and 3, respectively.
At the left end point of the collective coordinate in each panel,
the configuration corresponds to the ``HF-like'' states.
See text for details.
}
 \label{occ_number}
\end{figure*}
\subsection{Non-integrable case (1): Three-level pairing model}
\label{sec:three-level-model}

In contrast to the two-level model,
the TDHFB for the three-level model is non integrable.
We set the parameters of the system as follows:
$\Omega_1=\Omega_2=\Omega_3=\Omega=8$,
$\epsilon_1=-\epsilon_0$, $\epsilon_2=0$, $\epsilon_3=1.5\epsilon_0$,
and $g=0.2\epsilon_0$.
We use the parameter $\epsilon_0$ as the unit energy.
For the sub-shell closed configuration of $N=2\Omega=16$,
the HFB ground state changes from the normal phase to
the superfluid phase at $g_c=0.058\epsilon_0$.
We calculate a chain of systems with even particle numbers
from $N=14$ to $N=24$. 

We obtain three eigen frequencies for the moving-frame QRPA equation, 
on the collective path (Fig. \ref{omega_sq}). 
First of all, we clearly identify the zero mode with $\omega^2=0$
everywhere along the collective path.
This means that the pair rotation is separated from the other
degrees of freedom in the ASCC.
The frequency could become imaginary ($\omega^2<0$).
Except for the case of sub-shell closure ($N=2\Omega_1=16$),
the frequency rapidly increases near the end points.
The end points are given by points where the search for the next point
on the collective path in Sec.~\ref{sec:ASCC} fails.

We choose the lowest frequency squared mode, except for the zero mode,
as a generator of the collective path ($q^1$).
Figure~\ref{occ_number} shows variation of the occupation probability
of each single-particle state, as functions of the collective
coordinate $q^1$ on the collective path.
The most striking feature is that the collective path terminates
with special configurations which are given by the integer number
of occupation.
This is the reason why the search for the collective path fails
at both the ends.
At the end points,
the occupation of the level 3 ($\epsilon_3$) vanishes, while
those of the levels 1 and 2 become either maximum or minimum.
The left end of each panel in Fig.~\ref{occ_number} corresponds to 
a kind of ``Hartree-Fock'' (HF) state which minimizes
the single-particle-energy sum, $\sum_\alpha 2j_\alpha \epsilon_\alpha$.
The pairing correlation is weakened in both ends of the
collective path.

\begin{figure}[htbp]
 \begin{center}
  \includegraphics[width=60mm,angle=-90]{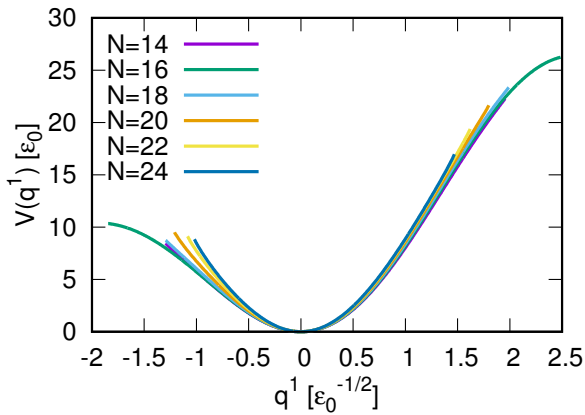}
 \end{center}
	\caption{Collective potential $V(q^1)$ obtained from the ASCC.
	We adjust the energy minimum point as $q^1=0$ and $V=0$.
}
 \label{potential}
\end{figure}

The collective mass with respect to the coordinate $q^1$ is normalized
to unity.
The collective potential is shown in Fig.~\ref{potential}.
The range of $q^1$ is the largest for the system with $N=16$.
This is because the variation of $j^1$ and $j^2$ is the largest in
this case.

\begin{table}[htbp]
\caption{Calculated excitation energies of the first and the second
excited states in units of $\epsilon_0$.
In the exact calculation, the second excited state in the ASCC+SPA
corresponds to the $0_4^+$ state.
See text for details.}
\label{ex}
\begin{ruledtabular}
\begin{tabular}{c|cccccc}
            $N$ & $14$ & $16$ & $18$ & $20$ & $22$ & $24$\\ \hline
ASCC+SPA (1st exc.) & $3.87$ & $3.90$ & $3.97$ & $4.09$ & $4.23$ & $4.33$\\
    Exact & $4.09$ & $4.13$ & $4.20$ & $4.30$ & $4.44$ & $4.60$\\ \hline
ASCC+SPA (2nd exc.)& $7.42$ & $7.42$ & $7.60$ & $7.92$ & $8.26$ & $8.47$\\
Exact & $7.65$ & $7.71$ & $7.88$ & $8.15$ & $8.49$ & $8.74$
\end{tabular}
\end{ruledtabular}
\end{table}

\begin{figure}[htbp]
 \begin{center}
  \includegraphics[width=60mm,angle=-90]{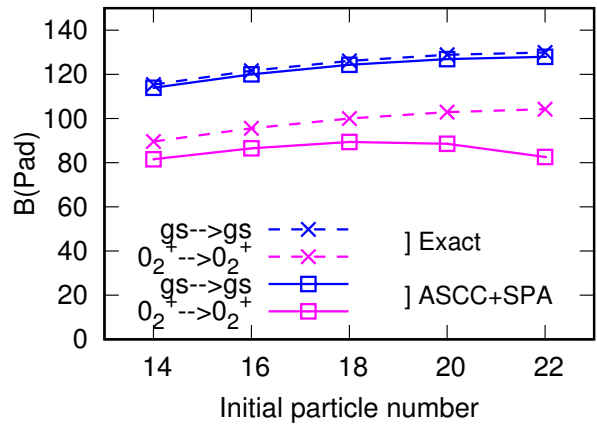}
 \end{center}
 \begin{center}
  \includegraphics[width=60mm,angle=-90]{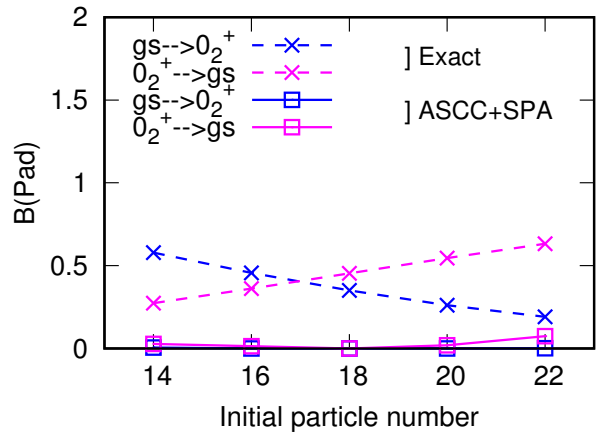}
 \end{center}
\caption{Calculated strength of pair-addition transition (\ref{BPad})
from $N=14$ to $N=22$. 
The solid (dashed) lines correspond to the ASCC+SPA (exact) calculation.
The horizontal line indicates the particle number of the initial states.
The upper panel shows the intraband transitions,
$0_1^+\to 0_1^+$ and $0_2^+\to 0_2^+$,
while the lower panel shows the interband transitions,
$0_1^+ \to 0_2^+$ and $0_2^+\to 0_1^+$.
}
 \label{3levelPad}
\end{figure}

Based on the collective path determined by the ASCC calculation,
we perform the requantization according to the SPA.
Table \ref{ex} shows the excitation energies of the first and second
excited states, determined by the EBK quantization condition (\ref{EBK}).
Comparing the result of the ASCC+SPA with that of the exact calculation,
we find that the excitation energies are reasonably well reproduced. 
The ASCC+SPA underestimates the excitation energies only by about 5 \%.

It should be noted that the second excited state in the collective path
corresponds to the $0_4^+$ state,
not to the $0_3^+$ state, in the exact calculation. 
We examine the interband $(k\neq k')$ pair-addition transition,
\begin{equation}
B(P_{\rm ad};k\rightarrow k') \equiv 
	\left|\bra{0_{k'}^+;N+2} \hat{S}^+ \ket{0_k^+;N}\right|^2 ,
\label{BPad}
\end{equation}
in the exact solution.
$B(P_{\rm ad};0_2^+\rightarrow 0_3^+)$ 
is $10\sim100$ times smaller than $B(P_{\rm ad};0_1^+\rightarrow 0_3^+)$,
while $B(P_{\rm ad};0_1^+\rightarrow 0_2^+)$ and
$B(P_{\rm ad};0_1^+\rightarrow 0_3^+)$ are in the same order.
The ASCC+SPA produces states in the same family,
namely, those belonging to the same collective subspace (path).
In the phonon-like picture,
we expect similar magnitude of the strengths for
$B(P_{\rm ad};{\rm g.s.}\rightarrow \omega_{\rm phon})$
and
$B(P_{\rm ad};\omega_{\rm phon}\rightarrow 2\omega_{\rm phon})$,
but smaller values of
$B(P_{\rm ad};{\rm g.s.}\rightarrow 2\omega_{\rm phon})$.
Thus, the $0_4^+$ state in the exact calculation
corresponds to the two-phonon state in the ASCC+SPA.
The $0_3^+$ state in the exact calculation may correspond to
a collective path associated with another solution of the moving-frame QRPA
(thin black line in Fig.~\ref{omega_sq}).


Next, we calculate the wave functions,
according to Eqs.~(\ref{SPA2}) and (\ref{SPA3}).
The ground state corresponds to the number-projected HFB state
(variation before projection).
In contrast, the excited states are given as superposition of
generalized Slater determinants in the collective subspace,
The pair-addition transition strengths
computed using these wave functions of the excited states
are shown in Fig. \ref{3levelPad}.
For the intraband transition ($k=k'$ in Eq. (\ref{BPad})),
the ASCC+SPA method well reproduces the strengths of the exact calculation.
The ground-to-ground transitions, $B(P_{\rm ad};0_1^+\rightarrow 0_1^+)$, 
are perfectly reproduced, while
$B(P_{\rm ad};0_2^+\rightarrow 0_2^+)$ are underestimated 
by about $10\%\sim20\%$.
\begin{figure*}[tb]
 \begin{center}
  \includegraphics[width=55mm,angle=-90]{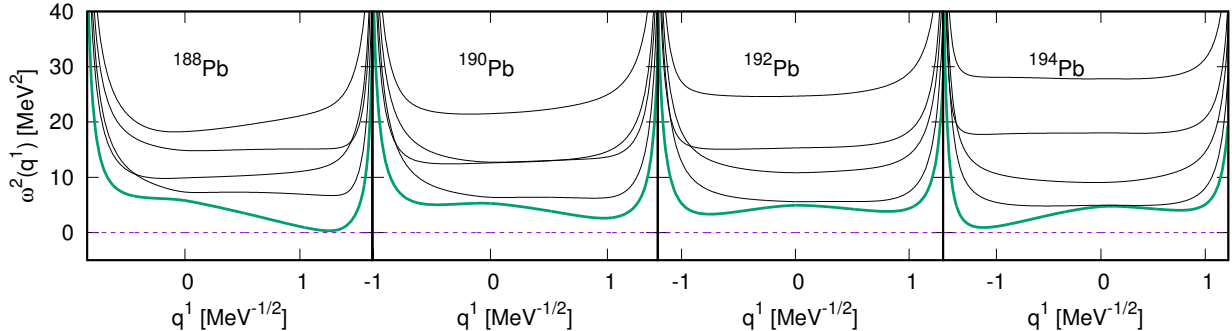}
 \end{center}
	\caption{The same as Fig. \ref{omega_sq} but for Pb isotopes.
}
 \label{Pb_omega_sq}
\end{figure*}

\begin{figure*}[tb]
 \begin{center}
  \includegraphics[width=55mm,angle=-90]{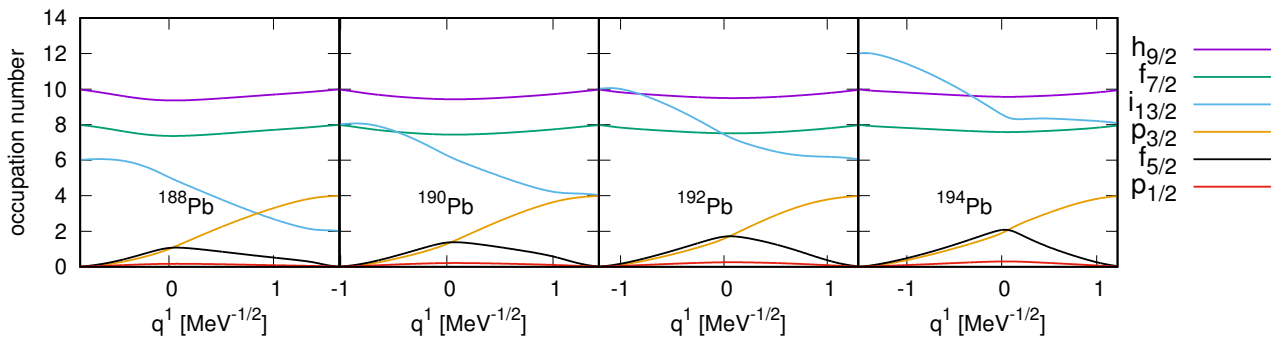}
 \end{center}
	\caption{The same as Fig. \ref{occ_number} but for Pb isotopes.
}
 \label{Pb_occ_number}
\end{figure*}

It is more difficult to reproduce the absolute magnitude of
interband transitions ($k\neq k'$), 
which are far smaller than the intraband transitions.
Although the increasing (decreasing) trend for
$B(P_{\rm ad};0_2^+\rightarrow 0_1^+)$
($B(P_{\rm ad};0_1^+\rightarrow 0_2^+)$)
as a function of the particle number is properly reproduced,
the absolute magnitude is significantly underestimated in the ASCC+SPA.
This is due to extremely small collectivity in the interband transitions.
Almost all the strengths are absorbed in the intraband transitions.
Even in the exact calculation, the pair addition strength is
about two orders of magnitude smaller than the intraband strength.
Remember that the non-collective limit ($g\rightarrow 0$) of this
value is $B(P_{\rm ad};0_1^+\rightarrow 0_2^+)=\Omega$.
Therefore, the pairing correlation hinders the interband transitions
by about one order of magnitude.
For such tiny quantities, perhaps, 
the reduction to the 1D collective path is not well justified.


We should remark here that there is a difficulty in the present ASCC+SPA
requantization for weak pairing cases.
In such cases, the potential minimum is close to the left end 
($q^1=q_L$) of the collective path, and the potential height
at $q^1=q_L$, $V(q_L) - V(0)$, becomes small.
Then, a classical trajectory with $E>V(q_L)$ hits this boundary ($q^1=q_L$).
In construction of wave functions,
the boundary condition at $q^1=q_L$ significantly influences the result.
In the present study, we choose a strong pairing case to avoid
such a situation.
As in Fig.~\ref{potential}, the potential height at $q^1=q_L$ has
about $10\epsilon_0$ which is larger than the excitation energies
of the second excitation.
Therefore, all the trajectories are ``closed'' in the usual sense.


\subsection{Non-integrable case (2): Pb isotopes}
\label{sec:Pb_isotopes}

Finally, we apply our method to neutrons' pairing dynamics in 
neutron-deficient Pb isotopes.
The spherical single-particle levels of neutrons
between the magic numbers 82 and 126 are adopted and
their energies are presented in Table \ref{Pb}. 
The coupling constant $g=0.138$ MeV is determined to reproduce
the experimental pairing gap given by the odd-even mass difference,
$\Delta(N)=\frac{(-1)^{N+1}}{2}[B(N+1)-2B(N)+B(N-1)]$ of ${}^{192}$Pb.
The even-even nuclei from ${}^{188}$Pb to ${}^{194}$Pb
are studied.
\begin{table}[tb]
\caption{Single-particle energies of Pb isotopes used in the calculation
in units of MeV.
These are obtained from the spherical Woods-Saxon potential
with the parameters of Ref.~\cite{BM69}.
}
\begin{ruledtabular}
\begin{tabular}{c|cccccc}
  orbit& $h_{9/2}$ & $f_{7/2}$ & $i_{13/2}$ & $p_{3/2}$ & $f_{5/2}$ & $p_{1/2}$\\ \hline
  energy(MeV)& $-10.94$ & $-10.69$ & $-8.74$ & $-8.44$ & $-8.16$ & $-7.45$\\
\end{tabular}
\end{ruledtabular}
\label{Pb}
\end{table}

The TDHFB dynamics is described by six degrees of freedom.
Figure \ref{Pb_omega_sq} shows eigenvalues of the
moving-frame QRPA equation.
Again, we find that there is a zero mode corresponding to
the neutron pair rotation.
Among the five vibrational modes, we choose the lowest one to
construct the collective path in the ASCC.
This lowest mode never crosses with other modes,
though the spacing between the lowest to the next lowest mode
can be very small, especially for ${}^{194}$Pb.
The evolution of the occupation numbers along the collective path
is shown in Fig. \ref{Pb_occ_number}.
Similarly to the three-level model, 
the end points of the collective path indicate exactly
the integer numbers, and
the left end of each panel corresponds to the
``Hartree-Fock''-like state.
On the collective path, the occupation numbers of $i_{13/2}$, $p_{3/2}$, and
$f_{5/2}$ mainly change.

\begin{figure}[bt]
 \begin{center}
  \includegraphics[width=60mm,angle=-90]{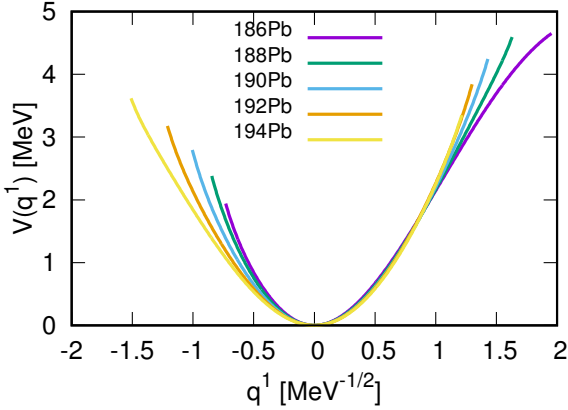}
 \end{center}
	\caption{The same as Fig. \ref{potential} but for Pb isotopes.
}
 \label{Pb_potential}
\end{figure}
The collective potentials for these isotopes are shown
in Fig. \ref{Pb_potential}.
The heights of the potentials at the left end, $V(q_L)-V(0)$,
are $2\sim3.5$ MeV.
For $^{186}$Pb, the height of the potential is not
enough to satisfy the condition, $E<V(q_L)$,
to have a closed trajectory for the first excited state
(See the last paragraph in Sec.~\ref{sec:three-level-model}).
We encounter another kind of problem for $^{196}$Pb,
which will be discussed in Sec.~\ref{sec5}.
Therefore, in this paper, we calculate the first excited states
in $^{188,190,192,194}$Pb.

We show the calculated excitation energy of the first excited state
in Table~\ref{Pb_ex}.
Experimentally, this pair vibrational excited $0^+$ state is
fragmented into several $0^+$ states due to other correlations,
such as quadrupole correlation, not taken into account in the present model.
We make a comparison with the exact solution of the multi-level
pairing model.
The ASCC+SPA method quantitatively reproduces the excitation energy of
the exact solution.

\begin{table}[bt]
\caption{The same as Table \ref{ex} but for Pb isotopes. The energies are given in units of MeV.}
  \begin{ruledtabular}
\begin{tabular}{c|cccccc}
   & ${}^{186}$Pb & ${}^{188}$Pb & ${}^{190}$Pb & ${}^{192}$Pb & ${}^{194}$Pb & ${}^{196}$Pb\\ \hline
ASCC+SPA & $-$ & $2.31$ & $2.21$ & $2.12$ & $2.04$ & $-$ \\ 
Exact & $2.58$ & $2.44$ & $2.34$ & $2.25$ & $2.20$ & $2.15$
\end{tabular}
\end{ruledtabular}
\label{Pb_ex}
\end{table}
\begin{figure}[htb]
 \begin{center}
  \includegraphics[width=60mm,angle=-90]{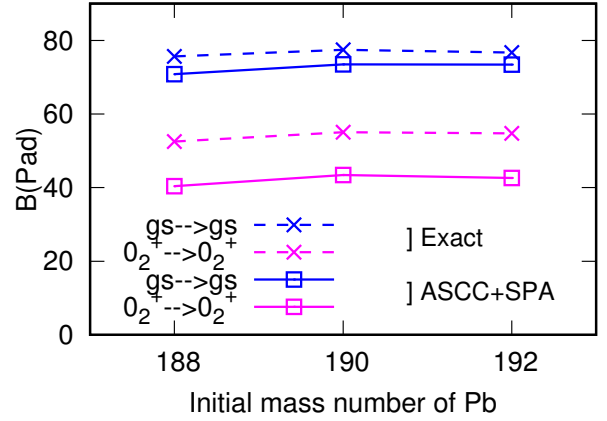}
 \end{center}
 \begin{center}
  \includegraphics[width=60mm,angle=-90]{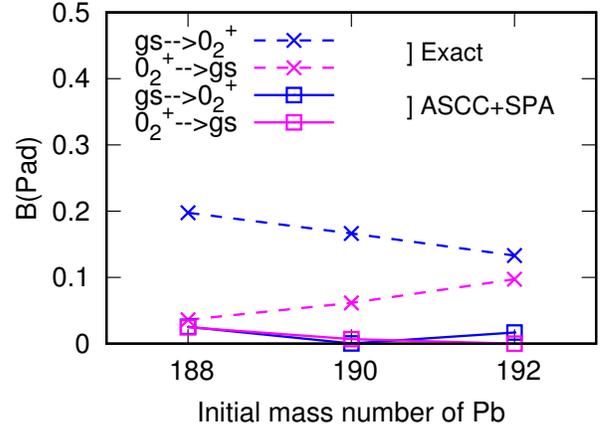}
 \end{center}
	\caption{The same as Fig. \ref{3levelPad} but for Pb isotopes.
}
 \label{PbPad}
\end{figure}
The pair-addition transition strengths are shown in Fig. \ref{PbPad}.
The feature that is similar to the three-level case is observed:
dominant intraband transition and very weak interband transitions.
The accuracy from the ASCC+SPA method well reproduces
$B(P_{\rm ad};0_1^+\rightarrow 0_1^+)$
and qualitatively reproduces
$B(P_{\rm ad};0_2^+\rightarrow 0_2^+)$ as well.
The deviation for the latter is about $25\%$.
The interband transitions are smaller than the intraband transitions
by more than two orders of magnitude.
This is also similar to the three-level model discussed
in Sec.~\ref{sec:three-level-model}.
For such weak transitions, the ASCC+SPA significantly
underestimates the strengths.
We may say that the ASCC+SPA gives reasonable results for the intraband
transitions in the realistic values of the pairing coupling constant $g$ and
single-particle levels.

\begin{figure}[t]
 \begin{center}
  \includegraphics[width=60mm,angle=-90]{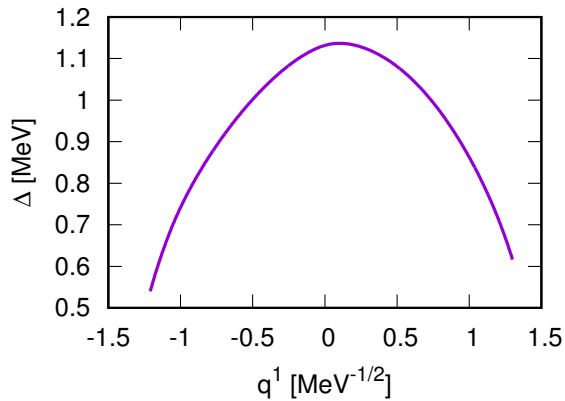}
 \end{center}
	\caption{Pairing gap as a function of collective coordinate $q^1$
	in ${}^{192}$Pb. 
}
 \label{192Pb_gap}
\end{figure}

Finally, we discuss the validity of the collective model approach
assuming the pairing gap as a collective coordinate.
The 5D collective Hamiltonian assuming the quadrupole deformation
parameters $\alpha_{2\mu} (\mu=\pm2,\pm1,$ and $0)$ as the collective coordinates
is widely utilized to analyze experimental data of quadrupole states.
Similarly, we may construct the pairing collective Hamiltonian
in terms of the pairing gap $\Delta$ and the gauge angle $\Phi$. 
As far as there is a one-to-one correspondence between $\Delta$ 
and the collective variable $q^1$ we obtained in the present study,
we can transform the collective Hamiltonian in $(q^1,\Phi)$
into the one in $(\Delta,\Phi)$.
The pairing gap $\Delta$ is defined as
\begin{align}
  \Delta (q) &\equiv \left. g\braket{\Phi,J;q,p|\hat{S}^-|\Phi,J;q,p} \right|_{\Phi=p=0} \nonumber \\
  &= g\sum_\alpha \sqrt{j^\alpha(\Omega_\alpha-j^\alpha)} .
\end{align}
Figure \ref{192Pb_gap} shows the pairing gap $\Delta$ in ${}^{192}$Pb
as a function of the collective coordinate $q^1$.
The peak in $\Delta$ is near $q^1=0$ and it is not a monotonic
function of $q^1$, thus no one-to-one correspondence exists.
The same behavior is observed for other Pb isotopes too.
Therefore, the pairing gap $\Delta$ is not a suitable collective coordinate
to describe the pairing dynamics in the multi-level model.



\section{Conclusion and discussion}
\label{sec5}

Extending our former work \cite{NN18},
which demonstrated the accuracy of the SPA for the requantization
of the TDHFB dynamics in the two-level pairing model, 
we propose the ASCC+SPA method for non-integrable systems.
In this approach, we use the ASCC method to extract the 2D
collective subspace including the pair rotation.
In other words, we extract an approximate integrable system 
in the non-integrable system described by $(q^1,p_1;J,\Phi)$.


We apply the ASCC+SPA method to the multi-level pairing model.
We investigate the three-level model and 
the multi-level model simulating Pb isotopes
with a realistic pairing coupling constant $g$ and single-particle levels.
In both cases, the low-lying excited $0^+$ states obtained with
the ASCC+SPA well reproduce the exact solutions
not only of the excitation energies but also of the wave functions.
In the ASCC+SPA, the pair-transition calculation is straightforward,
because we have a microscopic wave function for every quantized state.
This overcomes a disadvantage in the conventional canonical requantization
in which we need to construct a pair-transition operator
in terms of the collective variables only.

Although the overall agreement between the ASCC+SPA and the exact
calculations is good in general,
we have encountered several problems remaining to be solved.
First, we can calculate a classical trajectory bound by the
pocket of a potential. However, it is not trivial how to treat
``unbound'' trajectories that hit the end point of the collective path.
See the potentials in Figs. \ref{potential} and \ref{Pb_potential}.
This happens in the calculation of ${}^{186}$Pb
(Sec.~\ref{sec:Pb_isotopes}).
Probably, it is necessary to find a proper boundary condition
in the collective subspace.
For instance, the 5D quadrupole collective model has
such boundary conditions imposed by the symmetry property of
the quadrupole degrees of freedom \cite{KB67}.

The second problem occurred in the calculation of $^{196}$Pb,
in which we have encountered complex eigenvalues and eigenvectors
of the moving-frame QRPA equation.
This happens at a point where the two eigen frequencies become
identical, $\omega_1^2=\omega_2^2$, namely at a crossing point.
We do not have a problem for the crossing between the pair rotational
mode and the other modes.
Currently, we do not know exactly when the complex solutions emerge.

Another problem we need to solve is a description of the
quantum tunneling.
The tunneling plays an essential role in
spontaneous fission, sub-barrier fusion reaction, and
shape coexistence phenomena \cite{NMMY16,WN16,WN17,HNMM07,HNMM08}.
In the present ASCC+SPA, the classical trajectory
cannot penetrate the potential barrier.
Since the ASCC is able to provide the 1D collective coordinate,
the imaginary-time TDHF is feasible and may be a solution
to this problem \cite{Neg82}.
These remaining issues in the ASCC+SPA approach should be
addressed in future.



\begin{acknowledgments}
This work is supported in part by JSPS KAKENHI Grants No. 18H01209, 17H05194, and 16K17680,
and
by JSPS-NSFC Bilateral Program for Joint Research Project
on Nuclear mass and life for unravelling mysteries of r-process.
Numerical calculations were performed in part using COMA at the CCS,
University of Tsukuba.
\end{acknowledgments}

\bibliographystyle{apsrev4-1}
\bibliography{ref}

\end{document}